\documentclass[preprint,12pt]{elsarticle}

\usepackage{graphicx}
\usepackage{amssymb}

\usepackage{lineno}
\modulolinenumbers[5]
\def\beq#1\eeq{\begin{linenomath*}\begin{equation}#1\end{equation}\end{linenomath*}}

\usepackage{amsmath}
\usepackage[table]{xcolor}
\usepackage{url}
\usepackage[
singlelinecheck=false 
]{caption}

\journal{Nucl. Instr. Meth. A}

\begin{document}

\begin{frontmatter}


\title{Accurate Taylor transfer maps for large aperture iron dominated magnets used in charged particle separators and spectrometers}



\author[add1,add2]{E. Kazantseva\corref{cor1}}
\ead{kazantseva@temf.tu-darmstadt.de}
\author[add1,add2]{O. Boine-Frankenheim}
\author[add2]{H. Weick}
\author[add3]{M. Berz}
\author[add3]{K. Makino}
\cortext[cor1]{Corresponding author}
\address[add1]{Technische Universit\"at Darmstadt, Schlossgartenstra\ss e 8, 64289 Darmstadt, Germany}
\address[add2]{GSI, Helmholtzzentrum f\"ur Schwerionenforschung GmbH, Planckstra\ss e 1, 64291~Darmstadt, Germany}
\address[add3]{Michigan State University, Department of Physics and Astronomy, Center for Dynamical Systems, 567 Wilson Road, East Lansing, MI 48824, USA}

\begin{abstract}
For high-resolution separators like the projected Super-FRS at FAIR, an adapted and accurate ion-optical model considering realistic B-dependent magnet parameters is crucial in achieving the desired parameters (e.g. resolution) and to enable a fast optimization.   
Starting from the magnetic field measurements and simulations, rigidity-dependent Taylor transfer maps are generated for the Super-FRS preseparator dipole magnets. The effects of the magnetic saturation in the steel yoke on the image aberrations are analyzed.
\end{abstract}

\end{frontmatter}


\section{Introduction}
\label{S:1}
The growing demand in a field of discovering and investigating rare isotopes by means of fragment separators yields challenging restrictions on future facilities. The main task of a fragment separator is an in-flight separation of many different species of nuclides, produced from a primary ion beam behind a target.
The Super-FRS (SFRS), an in-flight projectile fragment separator, being built for the FAIR project at GSI \cite{Geissel:2003lcy}, is an example of combining high flexibility with ambitious design parameters. 

Due to its high design momentum resolution together with large angular and momentum acceptance (horizontal angular acceptance $A_h=\pm 40~\text{mrad}$, vertical angular acceptance $A_v=\pm 20~\text{mrad}$, and momentum acceptance $\Delta p/p=\pm 2.5\%$) the dipole magnets of the SFRS have large usable apertures of 120\,cm$\times 14$\,cm for radiation resistant preseparator dipoles and 38\,cm$\times 14$\,cm for the superconducting main separator dipoles. The actual vertical air gap is in both cases as large as 19\,cm. The design range of the particle magnetic rigidity $B\rho$ of 2-20\,Tm requires the variation of the main dipole magnetic field $B_0$  from 0.15\,T to 1.6\,T and of the coil current $I$ from 60\,A to 643\,A.
In the upper third of the field range, magnetic saturation effects are significant, leading to local changes of the magnetic field strength $\vec{B}$ and the corresponding particle orbits.

For the SFRS and similar separators and spectrometers, where frequent changes of $B\rho$ during operation is required for tuning and selection of different nuclides, it is important to have a fast ion-optical model with good predictability, especially for investigations involving rare nuclei with low production rates at high $B\rho$. Thus, to maintain the predictability of the ion-optical codes it is important to consider magnetic saturation effects in the underlying model. 

\begin{figure}[t!]
\includegraphics[width=0.5\textwidth]{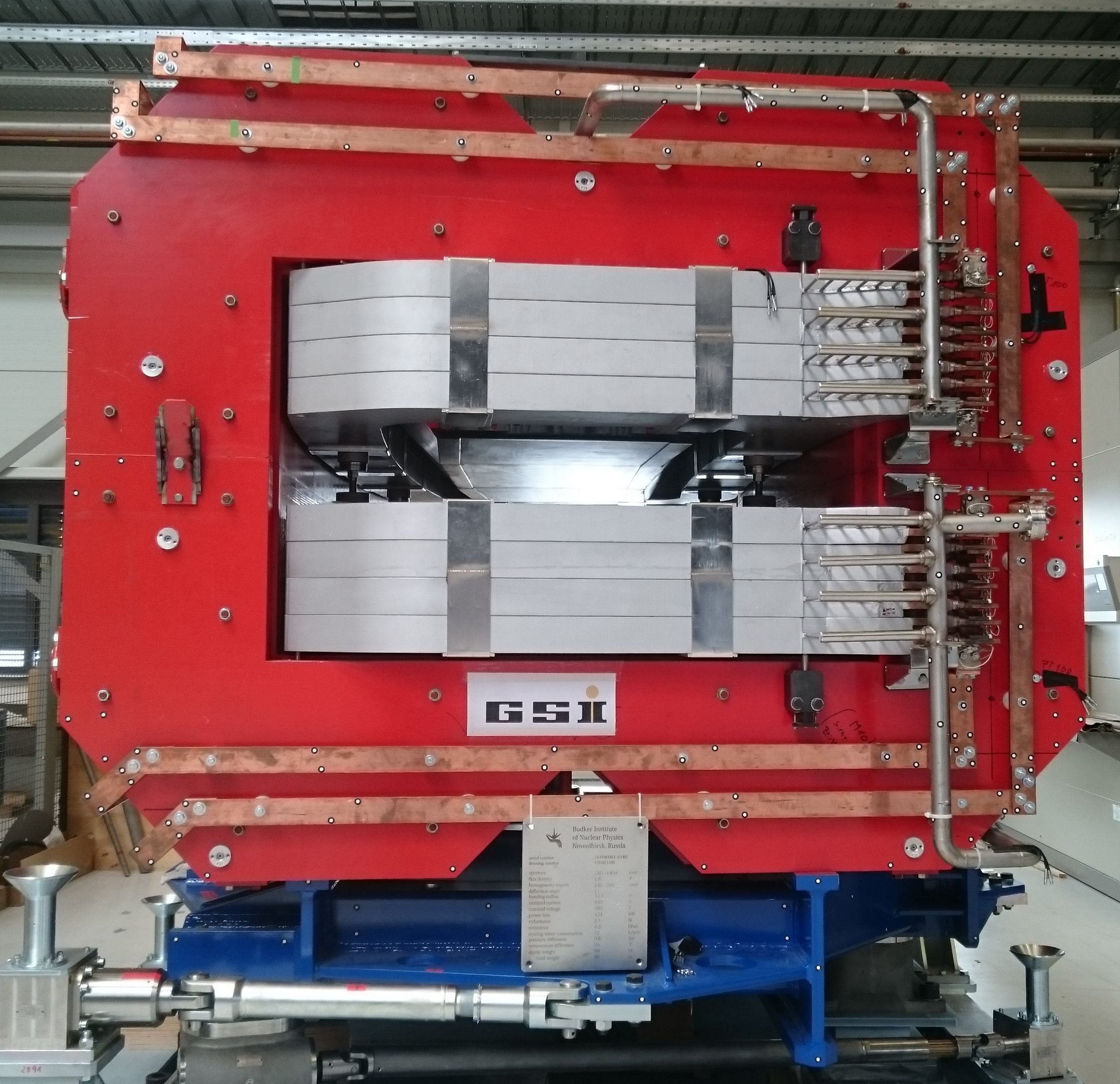}
\caption{The prototype of the normal conducting 11$^{\circ}$ Super-FRS dipole magnet with the design bending radius $R_0=12.5$\,m.}
\label{Fig:dip1}
\end{figure}

To obtain a fast and accurate ion-optical model for the SFRS, we have developed a general approach for polynomial representation of the magnetic field while exactly preserving its harmonic properties and computation of accurate transfer maps of arbitrary order starting from a 3D magnetic field distribution $\vec{B}(\vec{r},I)$. Here $\vec{r}=(X,Y,Z)$ is the position in the right-handed coordinate system of the magnet with the origin in the center, longitudinal direction $Z$ and vertical direction $Y$. This method is robust against the noisy data and allows for the use of measured magnetic field data as input.
The COSY INFINITY \cite{Makino:2006sx} and the Python programming language were used for the computations. The approach has been applied to the normal conducting radiation-resistant dipole magnet of the SFRS preseparator depicted in Fig. \ref{Fig:dip1} with design deflection angle $\theta_0=11^{\circ}$ and design radius $R_0=12.5$\,m \cite{Muehle}. The effects of the saturation are analyzed in detail in this paper.

\section{From magnetic field to transfer maps: a step by step description}
\label{S:2}
Our approach to compute realistic high order transfer maps can be divided into the following four steps:
\begin{enumerate}
\item Measurements or simulations of the magnetic field.
\item Determination of the reference trajectory.
\item Construction of the $\vec{B}$-field as smooth functions of coordinates $\vec{r}$ and excitation currents $I$.
\item Computation of transfer maps in the differential algebraic (DA) framework.
\end{enumerate}
The details of each step will be discussed in the following subsections for the example of the normal conducting SFRS-preseparator dipole magnet depicted in Fig. \ref{Fig:dip1}. All 3D simulations of the dipole were performed using the finite element method in the CST EMS magnetostatics solver \cite{CSTEMS}.

\subsection{Magnetic field measurements and simulations}
\label{S:MF}
In order to obtain a reliable transfer map of an ion-optical element, accurate magnetic field information from measurements or simulations is crucial in the complete region of the usable aperture. However, the magnetization of ferromagnetics used for most accelerator magnets is a complicated stochastic hysteretic process with a nonlinear dependency on a variety of parameters like the magnetic field strength, ramping rate, mechanical stress and temperature. This causes problems for both simulations and measurements.  

To partially resolve the hysteresis issue, the following rules for measurements and operation are commonly used in the accelerator community:
\begin{enumerate}
\item Only one hysteresis branch is used.
\item The ramp rate of the coil current is set slow enough to grant a quasi\-static behavior of the hysteresis curve.
\item The cycle of the magnetization is repeated until the resulting $B$ field becomes reproducible.
\item The coils are cooled to provide a stable operation temperature.
\end{enumerate}

Following these rules, the $\vec{B}(I)$ dependence becomes unambiguous down to the noise level (defined e.g. by the quality of the power supply) and it allows usage of simplified non-hysteretic simulation methods.
The most commonly used 3D magnetostatics simulation codes in the accelerator community such as CST EMS \cite{CSTEMS}, Opera Tosca \cite{Opera}, COMSOL AC/DC \cite{COMSOL} and Ansys Maxwell \cite{Ansys},
are based on approaches using the so-called virgin curve\footnote{Transient hysteresis simulation module available e.g. in Opera, is impractical for magnetostatic simulations due to much larger computational times.}.
The virgin $B$-$H$ curve starts at the point of the fully demagnetized state ($H$=0,$B$=0) and ends at one of the points of maximal absolute magnetization ($H_{\textrm{max}}$,$B_{\textrm{max}}$) or ($-H_{\textrm{max}}$,$-B_{\textrm{max}}$). This approach is sufficient for many applications with soft magnetic materials because of their narrow hysteresis curve.

\begin{figure}[t!]
\includegraphics[width=0.5\textwidth]{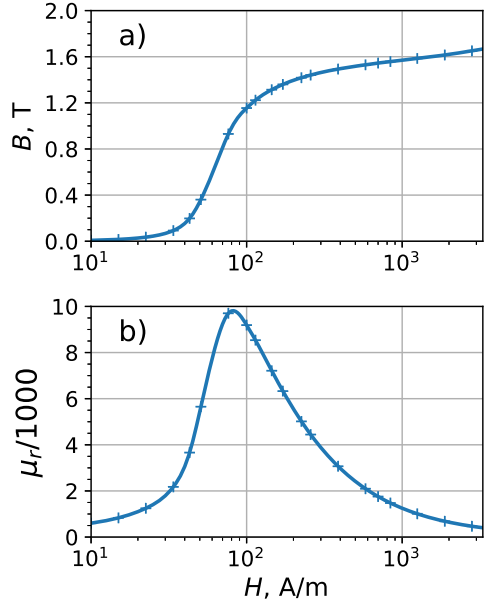}
\caption{The magnetization curve of the yoke steel \textbf{a)} and the corresponding relative permeability $\mu_r$ \textbf{b)} dependent on the magnetic field strength $H$. 
}
\label{Fig:bh}
\end{figure}

For the yoke material of the considered dipole the virgin curve was measured using a permeameter \cite{Henrichsen} at GSI. The measurement result and the corresponding relative permeability are plotted in Fig.~\ref{Fig:bh}.
The dipole will be powered by a unipolar current source. In operation the magnetization follows sub-branches starting from the remanent field $\vec{B}_r(\vec{r})$. Therefore, a considerable difference between the magnetic measurements and simulations can be expected for low currents.

\begin{figure}[t!]
\includegraphics[width=0.5\textwidth]{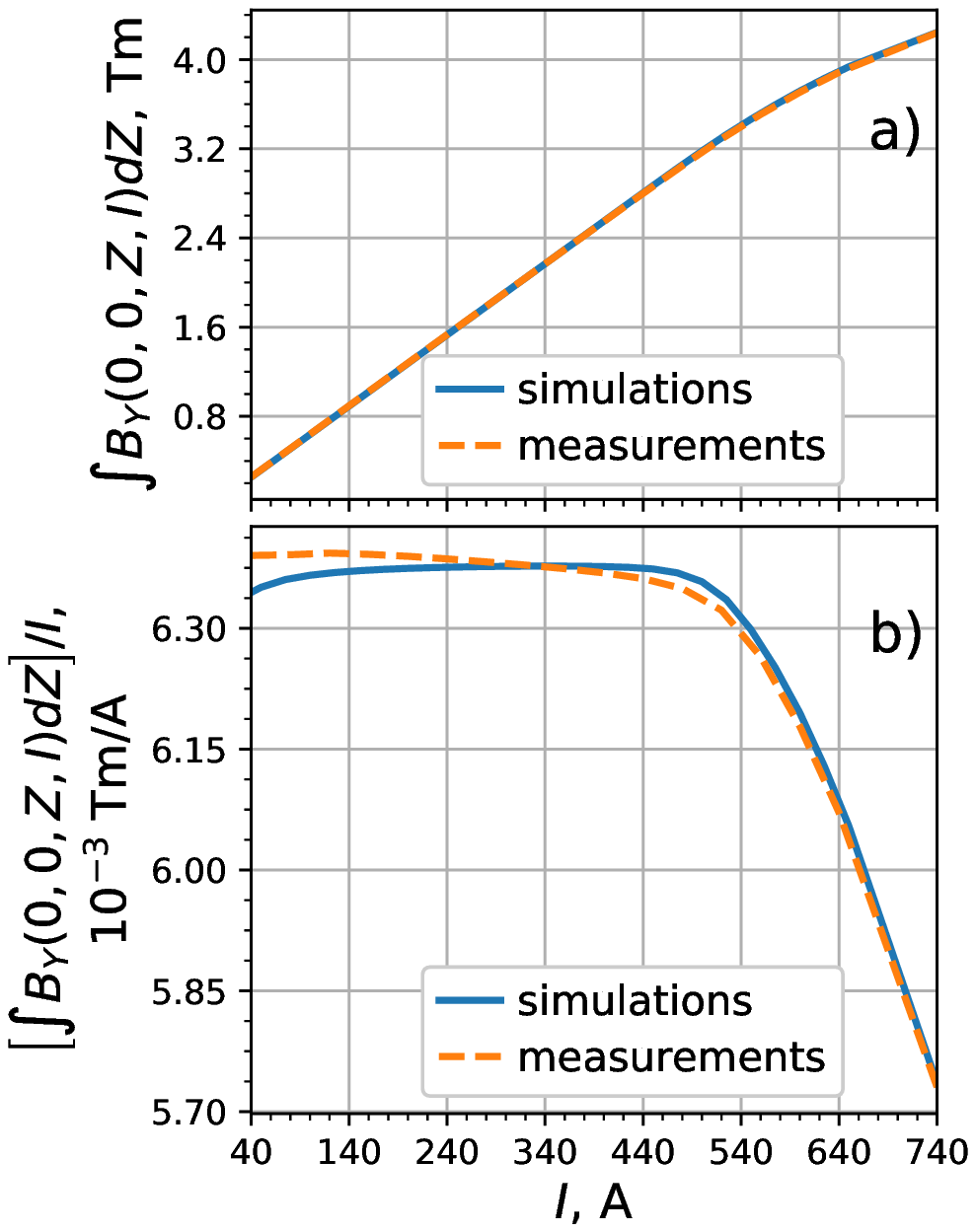}
\caption{Integral excitation curve $\int B_Y(0,0,Z,I)dZ$ \textbf{a)} and normalized integral excitation curve $\int B_Y(0,0,Z,I) dZ/I$ \textbf{b)} derived from simulations and measurements.}
\label{Fig:EC}
\end{figure}
The measured and simulated integral excitation curves are shown in Fig.~\ref{Fig:EC}~\textbf{a)} and appear nearly identical. Only after the normalization to $I$ the expected deviation of the simulated to the measured data is visible as depicted in Fig.~\ref{Fig:EC}~\textbf{b)}. The difference is maximal ($\approx0.8\%$) for $I=40$\,A and is significantly lower for higher currents. The slight shape deviation of the curves originates from the difference in the real and simulated magnetization processes. 

\begin{figure}[t!]
\includegraphics[width=0.6\textwidth]{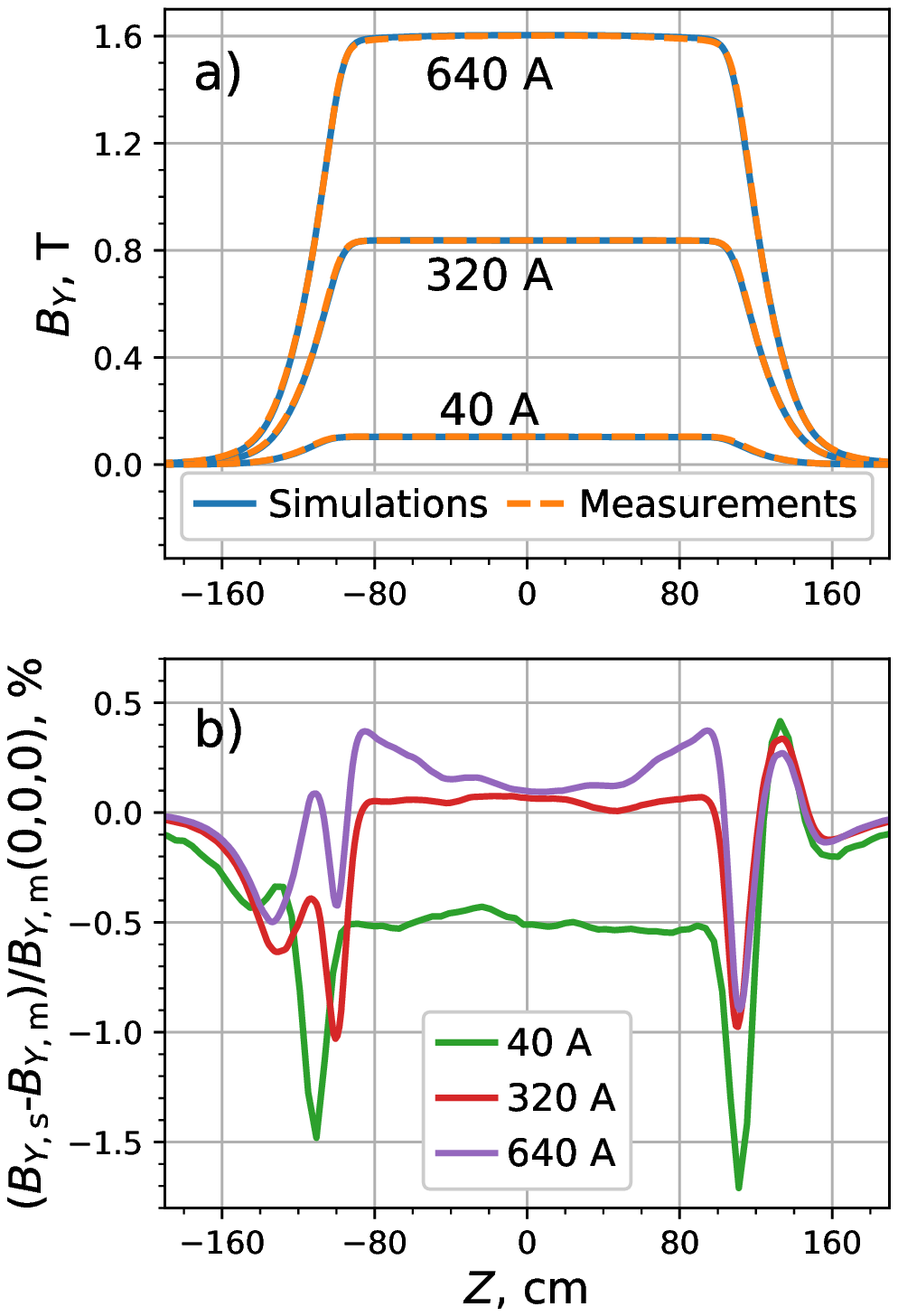}
\caption{Simulated $B_{Y,\textrm{s}}$ and measured $B_{Y,\textrm{m}}$ magnetic field along the $Z$ axis \textbf{a)} and relative error $(B_{Y,\textrm{m}}-B_{Y,\textrm{s}})/B_{Y,\textrm{m}}(0,0,0)$ \textbf{b)} for excitation currents of 40\,A, 320\,A and 640\,A.}
\label{Fig:ByzEr}
\end{figure}

The distributions of the $B$ field along the $Z$ axis for different $I$ values are depicted in Fig.~\ref{Fig:ByzEr}~\textbf{a)}. The relative difference between the measured $B_m$ and simulated $B_s$ field in Fig.~\ref{Fig:ByzEr}~\textbf{b)} in the main field region (-80\,cm to 80\,cm) originates mainly from the absense of the remanence in the simulations. The non-uniformity of the relative difference along the $Z$ axis can be explained by the different magnetization curves and unknown inhomogeneity of the magnetic properties of the yoke of the real dipole.

\begin{figure}[t!]
\includegraphics[width=0.5\textwidth]{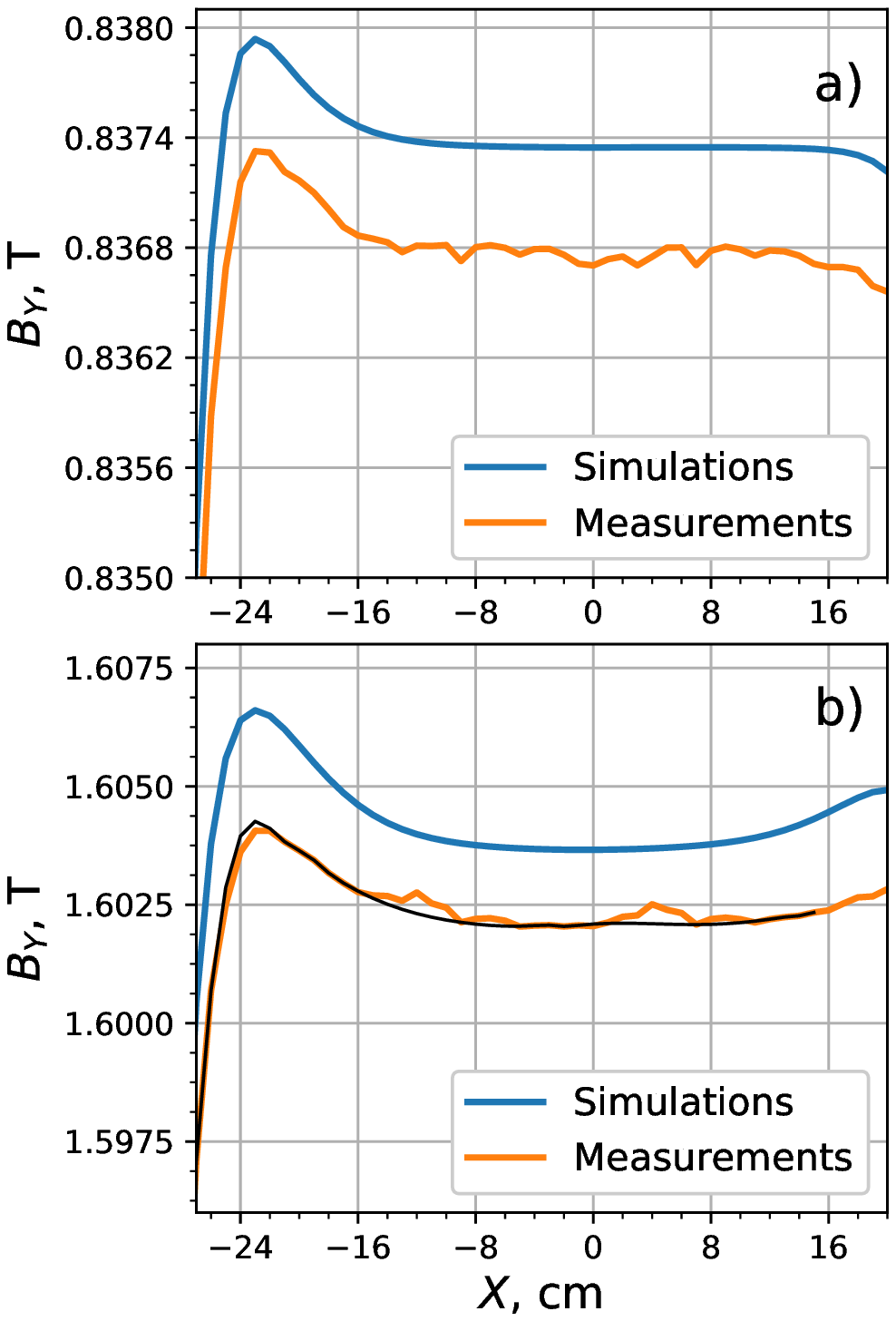}
\caption{Measured and simulated magnetic field along the $X$ axis (transverse direction) for coil currents of 320\,A \textbf{a)} and 640\,A \textbf{b)}. The ripples in measured data correspond to systematic measurement error. Dark-green line in \textbf{b)} is a result of removing the ripples from the measurement data.}
\label{Fig:Bxz}
\end{figure}

Nonetheless, despite deviations in the longitudinal distributions, the simulated transversal field distributions for higher currents are in good agreement with the measurements as shown in Fig.~\ref{Fig:Bxz} for 320\,A and 640\,A.  The ripples in measured data spread along the entire $Z$ axis and correspond to a systematic measurement error. After removing the ripples (black line in \textbf{b)}), the measured and simulated field distribution along the $X$ axis (transverse direction) in the middle of the magnet have a similar shape. 

\subsection{Setting the reference trajectory in dipole}
\label{S:refpathDip}
\begin{figure}[t!]
\includegraphics[width=0.5\textwidth]{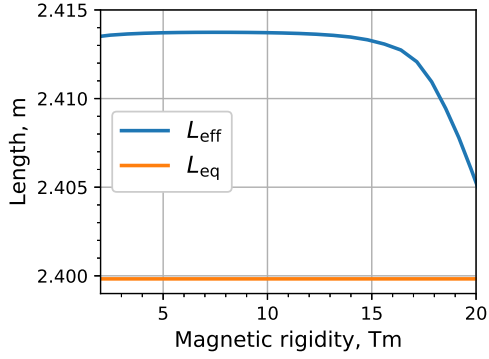}
\caption{Calculated effective $L_{\textrm{eff}}$ and equivalent $L_{\textrm{eq}}$ lengths versus the magnetic rigidity.}
\label{Fig:LeffvsLeq}
\end{figure}

\begin{figure}[t!]
\includegraphics[width=1\textwidth]{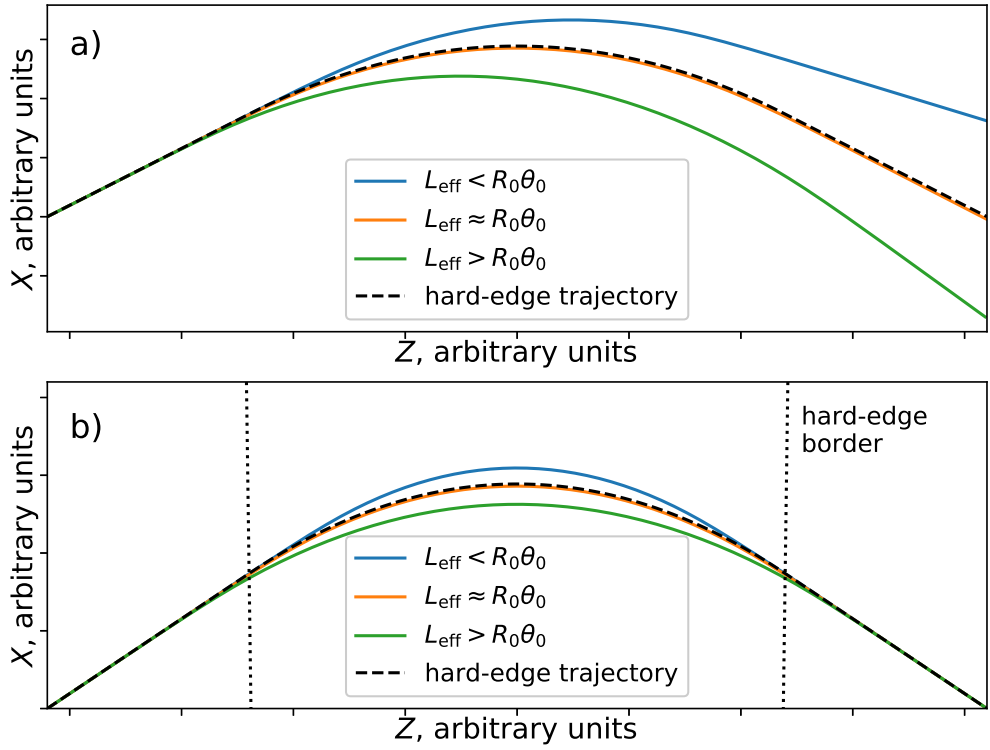}
\caption{Particle trajectories in sector dipoles with different effective lengths $L_{\textrm{eff}}$. In \textbf{a)} the trajectories are set so that $B\rho=B_0R_0$, whereas in \textbf{b)} $B\rho=\int_{-\infty}^{\infty}Bds/\theta_0$ is fulfilled.}
\label{Fig:RT_dif_Lef}
\end{figure}

An essential step for the ion-optical simulation of a dipole using measured or simulated fields is to set up the realistic reference trajectory in the magnet coordinates. This means to assign one of the possible realistic trajectories of a particle with central value of $B\rho$ as the reference. This trajectory should be located centrally in good field area and as close as possible to the ideal one. 
One issue which leads to changing of the particle trajectories with saturation of the yokes of the magnetic elements is the shortening of the effective length of a dipole
\beq\label{Eq:Leff}
L_{\mathrm{eff}}:=\frac{\int_{-\infty}^{\infty}B(S)dS}{B_{0}}.
\eeq
For the considered magnet, the $L_{\textrm{eff}}$ drops by 1\,cm between 2\,Tm and 20\,Tm as shown in Fig.~\ref{Fig:LeffvsLeq}.
Under the condition 
\beq\label{Eq:bhobrho}
B\rho=B_0\cdot R_0,
\eeq
the shortening of $L_{\textrm{eff}}$ leads to a decrease of the deflection angle of reference particle.
In general, changing the effective length while keeping $B_0=B\rho/R_0$ will lead to various deflecting angles, which might differ from the design value, as shown in the Fig.~\ref{Fig:RT_dif_Lef}~\textbf{a)}.
The situation can be improved using the equivalent (hard-edge) length 
\beq\label{Eq:Leq}
L_{\textrm{eq}}:=\frac{\int_{-\infty}^{\infty}B(S)dS}{B_{\textrm{eff}}}=\int_{-\infty}^{\infty}\frac{B(S)}{B\rho}dS\cdot R_0=\theta R_0,
\eeq
which is equal to the path arclength in a homogeneous sector magnet with a constant field \(B_{\textrm{eff}}=B\rho/R_0\), deflection radius $R_0$ and deflection angle $\theta$.
$L_{\textrm{eq}}$ was introduced as an alternative to $L_{\textrm{eff}}$ in \cite{Kazantseva:IPAC2017-WEPAB026}\footnote{In \cite{Kazantseva:IPAC2017-WEPAB026} the traditional effective length is named $L_{eff0}$ and the equivalent length is named $L_{\textrm{eff}}$.}. Unlike $L_{\textrm{eff}}$, $L_{\textrm{eq}}$ is an adjustable parameter, which can be set to a predefined value to achieve the design deflecting angle 
\beq
\theta_0 = \int_S \frac{B_Y(I)}{B\rho} ds
\label{Eq:setXi}
\eeq
when varying the coil current. In Fig.~\ref{Fig:LeffvsLeq} $L_{\textrm{eq}}(B\rho)$ is set to a constant value of $R_0\theta_0$. For different relations between effective and equivalent lengths this leads to a fixed deflection angle and slightly different curvatures for the reference particle as depicted in Fig.~\ref{Fig:RT_dif_Lef} \textbf{b)}.

For sector dipoles $L_{\textrm{eq}}$ can also be tuned by varying the position of the reference particle $X_{\textrm{i}}$ at the entrance of the magnet well outside of the field, where $B(X_{\textrm{i}},0,Z_{\textrm{i}})\approx 0$. This might be performed by shifting the dipole, which in general can be done only before the first operation of the machine. By tuning both $I$ and $X_{\textrm{i}}$, one can achieve $L_{\textrm{eff}}=L_{\textrm{eq}}$ at utmost two points\footnote{One for a monotone $L_{\textrm{eff}}(B\rho)$ and two, if it has a local extremum.} of $L_{\textrm{eff}}(B\rho)$ implying the simultaneous fulfillment of Eq.~(\ref{Eq:setXi}) and  
\beq
B_0(I)=B(X_0,Y=0,Z=0,I)=\frac{B\rho}{R_0},
\label{Eq:setI}
\eeq
where $X_0$ corresponds to the position of the reference trajectory in the middle (in $Z$-direction) of the magnet. 

\subsection{Calculation of the DA representation of the magnetic field}
\label{S:SIHM}
For large aperture machines with wide rigidity ranges it is advantageous to have a quick model of the detailed magnetic field information for any coil current. Therefore, a magnetic field distribution $\vec{B}(X,Y,Z,I)$ can be described using a set of 4-variable polynomials for its approximation. In addition, the polynomial field representation saves time for accessing and evaluation as well as storage space in comparison to usual 3D arrays of field values. 
These polynomials can be obtained with the surface integration Helmholtz method (SIHM) in the DA framework \cite{Manikonda2006a,Manikonda2006}. SIHM finds a harmonic vector field inside of a source-free simply-connected volume, given the vector field on the surface of the volume. In our case the volume was chosen as a rectangular box. The resulting magnetic field components are DA vectors, i.e., they represent the Taylor expansion coefficients of $B_{X,Y,Z}$ in $X$, $Y$ and $Z$ up to a predefined order \cite{Berz:1999we}. 
The integration over the surface in SIHM makes the method robust against the random input errors, e.g. measurement errors. Besides that, even if the errors in the initial magnetic field break its harmonic property, SIHM enforces that $\Delta \vec{B}=0$ holds for the resulting magnetic field up to machine accuracy.

To take the current dependency of the $B$ field into account, the field components can be decomposed into a superposition
\beq\label{eq:BI}
B_\alpha(I)\approx b^\alpha_0+b^\alpha_1(I-I_0)+b^\alpha_2(I-I_0)^2+...+b^\alpha_n(I-I_0)^n,
\eeq
where $\alpha\in \{X,Y,Z\}$ and $I_0$ is the expansion point in $I$. The coefficients $b^\alpha_{0,1,2,...,n}$ can be used as input for the SIHM procedure in COSY INFINITY. The output DA vectors can be recombined using Eq.~(\ref{eq:BI}) yielding $\vec{B}(X,Y,Z,I)$ inside the volume of interest with the DA variable $(I-I_0)$.
In case of midplane symmetry it is possible to obtain a 3D $\vec{B}$ distribution from a 2D $B_Y$ distribution in plane $Y=0$ using DA fixed point theorem\footnote{The method is available in the beam physics package of COSY INFINITY \cite{Makino:2011zz}.}, reducing the number of fitted coefficients $N_c$ from ${(n+3)!}/(n!\,3!)$ in 3D to ${(n+2)!}/(n!\,2!)$ in 2D.

Although SIHM computes the field and its derivatives very accurately, for relatively flat volumes (i.e. where one dimension is smaller than the others) it is not well suited for obtaining a high order polynomial, which would represent the field in the entire transversal cut of volume of interest.
This is due to the fact, that in DA SIHM routine, the integrand \(1/|\vec{r}_v-\vec{r}_s|\) \cite{Manikonda2006, Manikonda2006a} is expanded in a volume expansion point $\vec{r}_v$ and a surface expansion point $\vec{r}_s$. This integrand is not analytical for $\vec{r}_v=\vec{r}_s$. Thus, the Taylor expansions do not converge for \(|\vec{r}-\vec{r}_v|>|\vec{r}_v-\vec{r}_s|\).
For the SFRS preseparator dipole magnet the physical vertical aperture is 18\,cm, which is smaller than the used horizontal aperture of $\pm 19$\,cm \cite{ParListSFRS} and in order to obtain the Taylor polynomials, we combine SIHM with a least squares fit. The second order DA vectors of the $B$ field\footnote{First and mixed second order partial derivatives are still accurate for a small convergence radius.} were calculated in a set of points in plane $Y$=0 covering the area of interest. The least squares fit was used to obtain higher order polynomials in the set of points along the reference path. The polynomials describe the initial $B$ field in the whole transversal area of interest and in the longitudinal direction on a length equal to the vertical aperture. 

\begin{figure}[t!]
\includegraphics[width=1\textwidth]{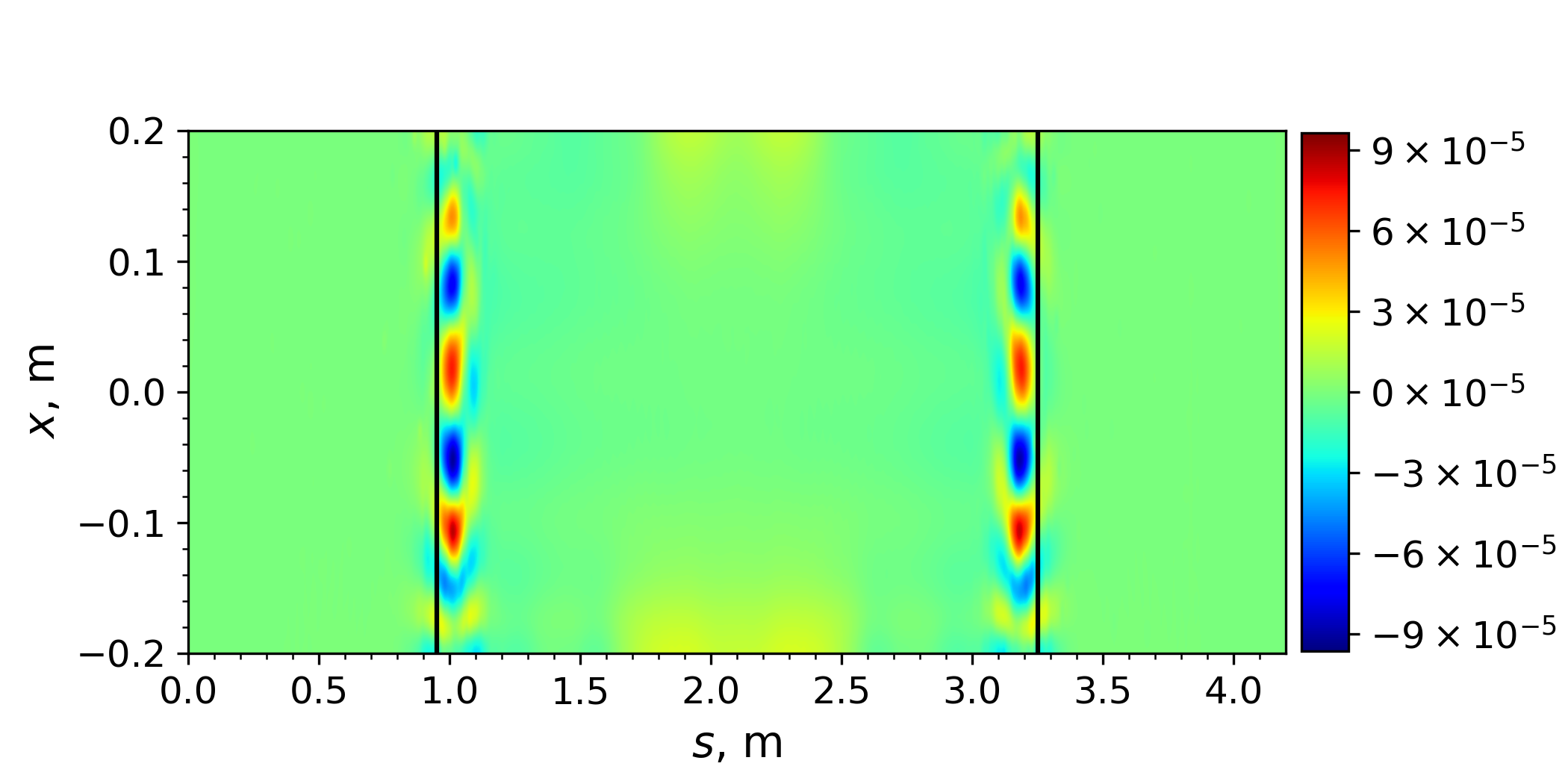}
\caption{Relative difference $\Delta B/B(0,0,0)$ between the initial magnetic field obtained from a FEM simulation and the 10\textsuperscript{th} order polynomial approximations along the reference path in curvilinear coordinates for a coil current of 575\,A. Black lines indicate the physical borders of the dipole.}
\label{Fig:575er}
\end{figure}

Using the methods described in this section, we obtained the $\vec{B}(X,Y,Z,I)$ polynomials for the considered dipole. In Fig.~\ref{Fig:575er} the relative error of the resulting field is depicted in the area of interest along the reference path, where 10\textsuperscript{th} order polynomials are used for the approximation and the coil current is 575\,A. The resulting polynomials are in a good agreement with initial field from FEM simulation. The highest error values are located on the fringes, where the field changes rapidly.

\begin{figure}[t!]
\includegraphics[width=0.5\textwidth]{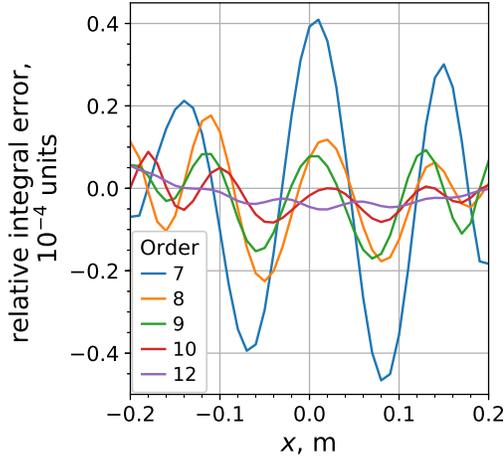}
\caption{Relative integral error $\int\left( B_Y(x,s)-B_{Y0}(x,s)\right) ds/\int B_{Y0}(x,s)ds \cdot 10^4$ with the initial magnetic field from a FEM simulation $B_{Y0}$ and the  magnetic field from polynomial approximations $B_{Y}$ calculated for different orders against the transverse curvilinear coordinate $x$ for a coil current of 575\,A. The integration is performed along the path length $s$.}
\label{Fig:575inter}
\end{figure}
In  Fig. \ref{Fig:575inter} the relative error integrated along the reference path is shown for different orders of the $B(X,Y,Z)$ approximation. The error originates mostly from the fringe field region and is oscillating along the transverse $x$-axis, where the oscillation amplitude decreases as the order increases. The non-zero mean value of the error results from the limited accuracy of the $B$-$I$-polynomial approximation and the accuracy of FEM-simulations (10$^{-6}$ in our case).
Practically, the choice of the order of approximation requires insight into the beam dynamics of the particular application.

\subsection{Finding the optimal current and obtaining Taylor transfer maps}
\label{S:5}
Using polynomial representation of the magnetic field $B(X,Y,Z,I)$ it is possible to obtain transfer maps for any required rigidity. We used two different methods of the transfer map calculation: one general method, and one method which treats the main field region and fringe fields separately.
The general method is based on application of the 8\textsuperscript{th} Runge-Kutta DA integrator in COSY INFINITY \cite{Makino2002} on a set of canonical beam physics equations of motion \cite{berz2014introduction}. In this paper we denote such maps as ``3D maps". 
With the other method, the maps are calculated using thick multipoles for transversal non-uniformities using the procedure MS (an inhomogeneous combined function bending magnet) in COSY INFINITY together with the Enge-function approximation for the fringe fields (``MS + Enge FF").
Both methods require the knowledge of the relation between the coil current and the magnetic rigidity, which we obtained as follows.
The function \[I(B\rho)=C_0^I+C_1^I(B\rho-B\rho_0)+C_2^I(B\rho-B\rho_0)^2+...C_N^I(B\rho-B\rho_0)^N\] should provide a correct deflection angle, which reduces to an optimization problem $\theta \overset{!}{=} \theta_0$. We used the shooting method to solve this problem. Due to the orthogonality of different order monomials in the DA framework, the coefficients $C_{\textrm{i}}^I$ can be fitted individually starting with $C_0$ and ending with $C_N$.

The 3D maps can be computed using $B(X,Y,Z,I)$ and taking $I(B\rho)$ into account directly in the equations of motion resulting in $B\rho$-dependent transfer maps. 

For the MS+Enge FF maps, the integral field harmonics and Enge coefficients were evaluated for a set of rigidity values.
\section{Application: Super-FRS preseparator optics with high order $B\rho$-dependent maps}
\label{S:6}
\begin{figure}[t!]
\includegraphics[width=1\textwidth]{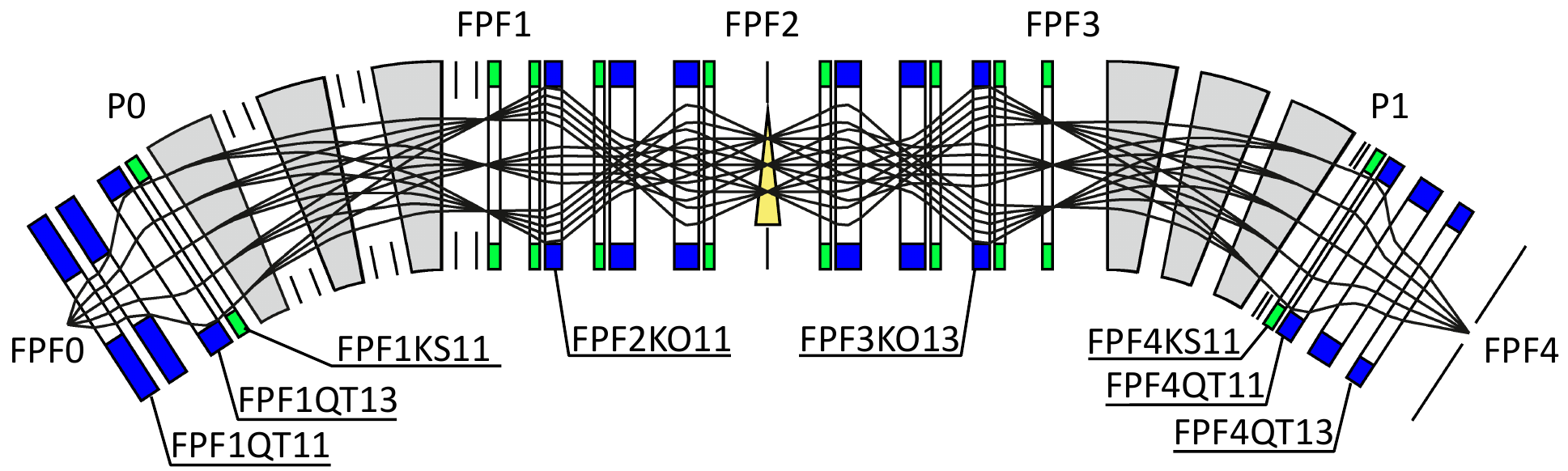}
\caption{Ion-optical layout of the SFRS preseparator with production target at FPF0 and focal planes FPF1-4. The wedge energy degrader can be placed in the FPF2 plane. The gray sectors denote dipole magnets, blue-marked elements denote quadrupoles (sometimes with octupole correctors), and light green-marked elements denote sextupoles. The rays produced for 5 initial angles and 3 initial energies demonstrate the achromatic layout of the SFRS-preseparator. More labels are used for further discussion in this paper.}
\label{Fig:Presep}
\end{figure}
The SFRS preseparator is a $B\rho - \Delta E - B\rho$ separator with two deflecting stages and a wedge energy degrader between them with its layout shown in Fig. \ref{Fig:Presep}. Each deflecting stage has a triplet of similar 11$^\circ$ dipoles. There are four focal planes in the SFRS preseparator. The most interesting planes are the dispersive focal plane FPF2, where the degrader is placed, and the achromatic focal plane FPF4 at the end. The detailed description of the ion-optical layout of the preseparator can be found in \cite{GEISSEL2006368}. 
Quadrupoles, sextupoles, and octupoles are used for focusing and correction of geometric and chromatic aberrations.

To study the impact of the $B\rho$-dependency and high order aberrations on the resolution of the SFRS preseparator the maps of the dipoles obtained in this work were inserted into the ion-optical model in COSY INFINITY. Within this section only 3D maps were used.
For the multipole elements we used standard COSY Enge fringe fields in this study.

There are two modes of the operation of the SFRS preseparator: separator mode and spectrometer mode. In the separator mode, the full layout of the preseparator is achromatic for the nuclei to be selected (see rays in the Fig. \ref{Fig:Presep}). The wedge degrader, placed in the dispersive FPF2, reduces the energy of the nuclei depending on their atomic number and hence grants the spatial separation of the nuclei with different atomic numbers at the FPF4.

For a successful operation of the SFRS in the separator mode, the beam has to be centralized for all rigidities. Therefore, the reference path in dipoles needs to be set up in a way to preserve the deflection angle, as described in subsection \ref{S:refpathDip}. Otherwise changing $L_{\textrm{eff}}$ would lead to shifting of the horizontal position of the beam by about 1 cm at the FPF2. 

For the best separation, the beam spot at FPF4 has to be minimized, which can be performed by reducing first and second order geometric (primarily horizontal) and chromatic aberrations in the focal planes via fitting of the multipole strengths. We tuned all multipoles of the preseparator (12 quadrupoles, 10 sextupoles and 4 octupoles) to achieve optimal settings for the rigidity range from 2 to 20 Tm and to preserve the first order ion-optical layout described in \cite{GEISSEL2006368}. 
The horizontal beam width inside the dipoles was kept constant, preserving the first order resolving power at FPF2: 
\beq\label{Eq:Res1}
R_{1,\mathrm{FPF2}}=|(x|\delta)|/((x|x)\Delta x_{\textrm{i}})\approx 2.6/(1.65\Delta x_{\textrm{i}}),
\eeq
which corresponds to $p/\Delta p \approx 1576$ for $\Delta x_{\textrm{i}}=1$ mm. 
The other fit conditions were to improve the preseparator transmission or to limit the multipole coil currents for the high rigidities. The same fit conditions were used for all rigidities.

\begin{figure}[t!]
\includegraphics[width=1\textwidth]{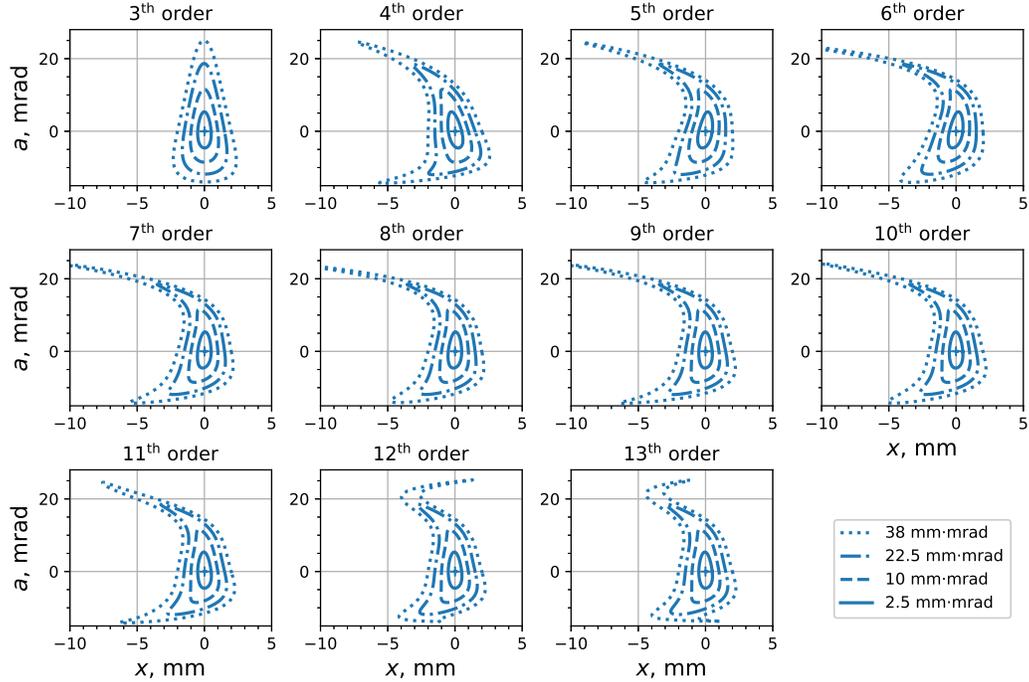}
\caption{Horizontal phase space images at FPF4 for different order transfer maps. Initial coordinates are given by functions ($x_k\cdot \cos(\phi),a_k\cdot \sin(\phi),0,0,0$) with $\phi \in [0,2\pi)$, $x_k \in$ \{0.25\,mm, 0.5\,mm, 0.75\,mm, 1\,mm\} and $a_k \in$ \{10\,mrad, 20\,mrad, 30\,mrad, 38\,mrad\}.}
\label{Fig:xaFPF4orders}
\end{figure}

Before proceeding with ion-optical studies, it is important to know the polynomial order of the transfer maps of the dipole that is sufficient for the SFRS application. Therefore we compare the horizontal phase space images at FPF4 for different orders and the same initial coordinates, as shown in Fig. \ref{Fig:xaFPF4orders}. For simplicity we consider the image aberrations of particles with initial distributions laying on concentric ellipses in horizontal phase space. Fig \ref{Fig:xaFPF4orders} shows that for an emittance of 22.5\,mm$\cdot$mrad, the resulting image does not change significantly beyond the 7\textsuperscript{th} order. For an emittance of 38\,mm$\cdot$mrad, corresponding to the maximal acceptance of the SFRS, the image stabilizes only after the 12\textsuperscript{th} order, since the lower orders display incorrect behavior of the top and bottom ends of the final phase space. 

\begin{figure}[t!]
\begin{minipage}{0.42\textwidth}
\includegraphics[width=1\textwidth]{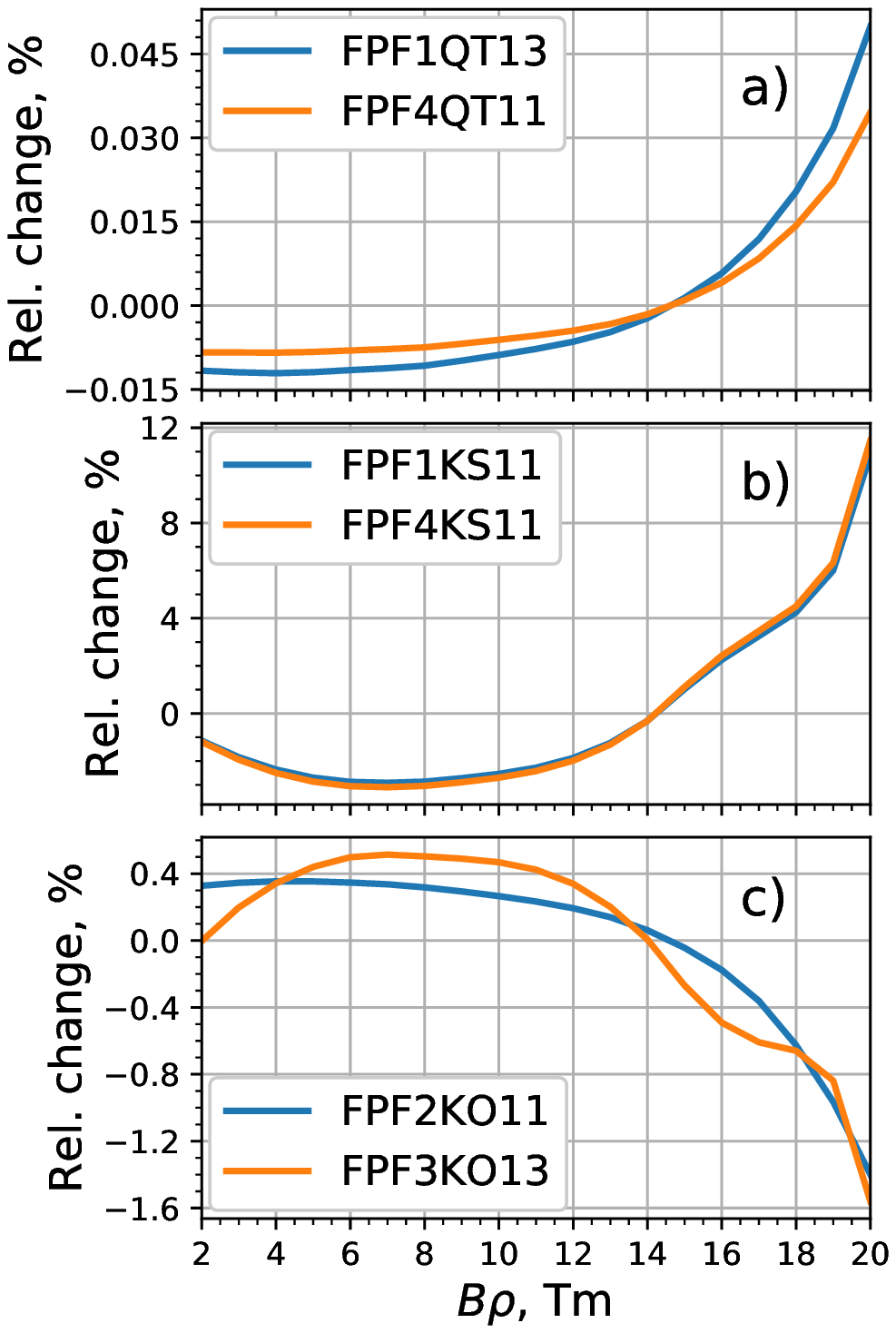}
\caption{Relative optimal multipole strengths versus the particle rigidity $B\rho$ for two quadrupoles \textbf{a)}, two sextupoles \textbf{b)} and two octupoles \textbf{c)} labeled in Fig.~\ref{Fig:Presep}.}
\label{Fig:PRESEPRelChQSO}
\end{minipage}\hspace{0.05\textwidth}
\begin{minipage}{0.44\textwidth}
\includegraphics[width=1\textwidth]{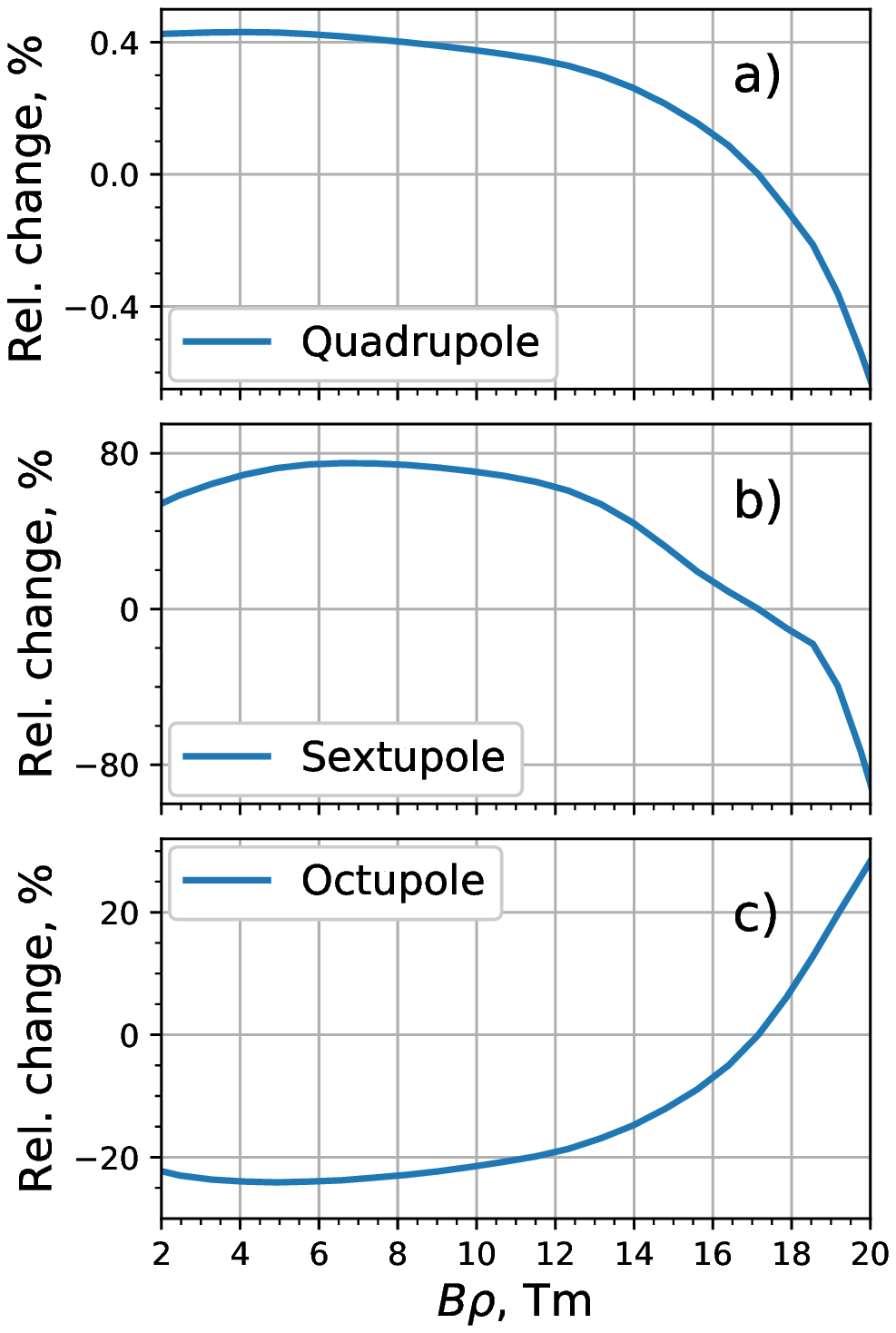}
\caption{Normalized relative integral non-uniformities of \textbf{a)} 1\textsuperscript{st} (quadrupole), \textbf{b)} 2\textsuperscript{nd} (sextupole) and \textbf{c)} 3\textsuperscript{rd} (octupole) orders versus the particle rigidity $B\rho$.}
\label{Fig:DipRelChQSO}
\end{minipage}
\end{figure}

Using the 12\textsuperscript{th} order dipole transfer maps we found the optimal multipole settings for the rigidity range 2-20 Tm using the multiparametric fit-procedure in COSY INFINITY. The relative change of the optimal multipole settings in preseparator optics (Fig. \ref{Fig:PRESEPRelChQSO}) have shapes very similar to the changes of corresponding integral non-uniformities in the dipole magnet field distribution (Fig. \ref{Fig:DipRelChQSO}), although with a different sign to compensate for the effect from the dipole. The magnets chosen for comparison are labeled in Fig. \ref{Fig:Presep}. The curve for octupole FPF3KO13 in Fig. \ref{Fig:PRESEPRelChQSO} has another shape which is likely influenced by the vertical octupole component of the dipole and corresponding fit conditions.

\begin{figure}[t!]
\includegraphics[width=0.6\textwidth]{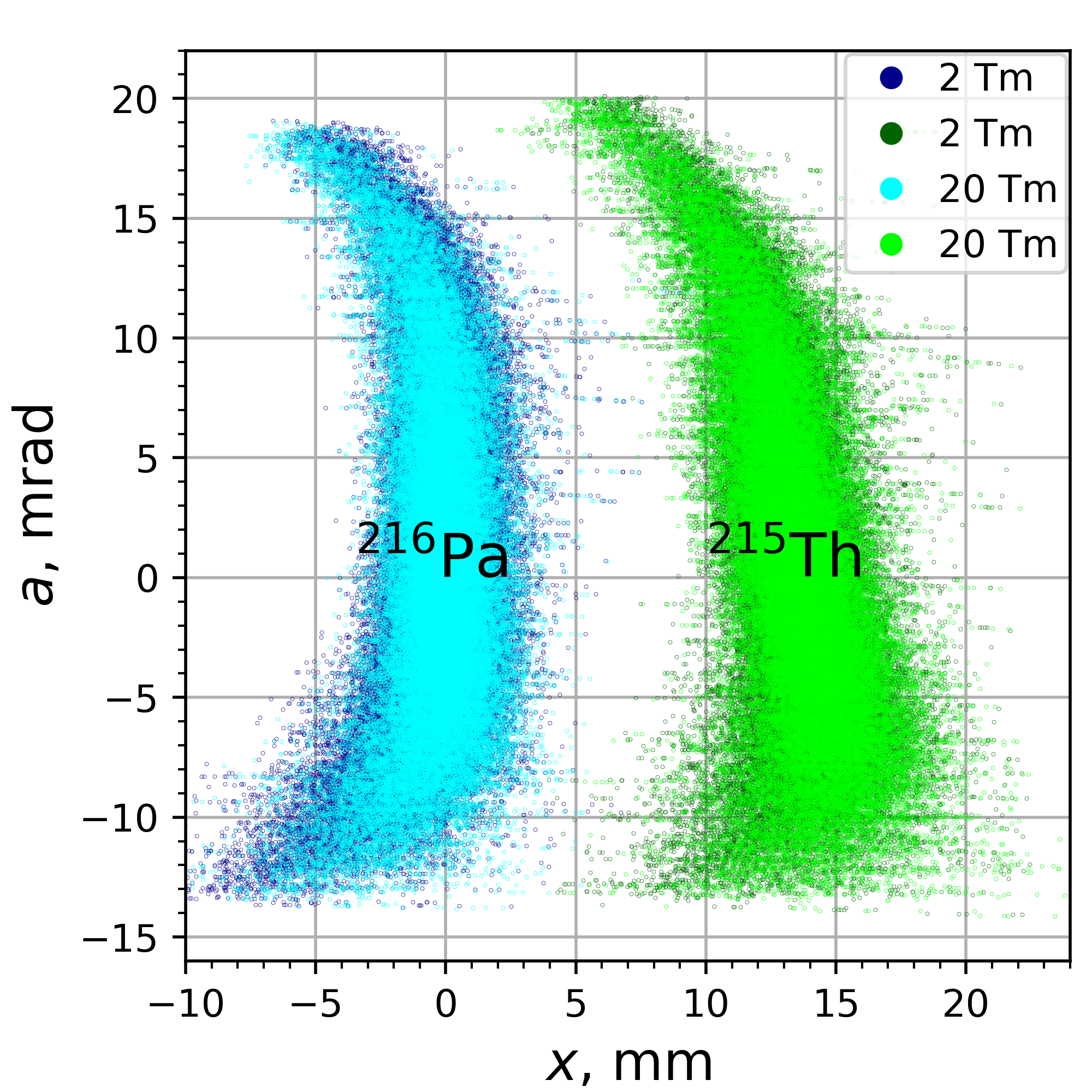}
\caption{Horizontal phase space images of the separation of fully-stripped 20 Tm $^{216}$Pa and $^{215}$Th after the preseparator including the copper wedge degrader, which slows the reference particle ($^{216}$Pa) down to 14 Tm. In the preseparator optics the transfer maps of the dipole magnets for 2\,Tm (dark dots) and 20\,Tm (light dots) were used to identify the maximal effect of the magnetic saturation on the separation.}
\label{Fig:FPF4_sep}
\end{figure}

To study possible changes that the magnetic saturation introduces into the predicted separation, we performed a numerical experiment by tracking two fragments of $^{238}$U from a carbon target, namely $^{216}$Pa and $^{215}$Th with rigidity 20 Tm, through the preseparator including the energy loss in the copper wedge degrader, slowing the reference particle ($^{216}$Pa) down to 14 Tm. In this case the resolution is limited by the inevitable energy loss straggling in the degrader, which was taken into account using the theory in \cite{Lindhard:1996zz}. The average energy loss was calculated using the Bethe-Bloch formula. For the computational convenience, all tracking simulations were performed using the rigidity of 20 Tm, whereas the energy/momentum deviations were scaled appropriately. The transfer maps were also scaled using the SYSCA method in COSY INFINITY \cite{Hoffstaetter:1996yt}. The particles that exceeded the local acceptance in the phase space were excluded from further tracking.
To observe the maximal possible change in the separation caused by the magnetic saturation, we compared the images on the horizontal phase space, which are produced with transfer maps for 2 and 20 Tm. The resulting phase space distribution is shown in Fig. \ref{Fig:FPF4_sep}, where dark blue and dark green dots correspond to 2 Tm and light blue and light green dots correspond to 20 Tm. In both cases, the initial beam phase space after the production target was the same: ($x,a,y,b,\delta$)=$\pm$(0.5 mm, 38 mrad, 2 mm, 20 mrad, 2.5\%). Although the effect of the saturation on the images of the $^{216}$Pa and $^{215}$Th on the achromatic focal plane FPF4 can be distinguished in Fig. \ref{Fig:FPF4_sep}, its magnitude is small, thus it has no meaningful effect on resolution.  

\begin{figure}[t!]
\includegraphics[width=0.5\textwidth]{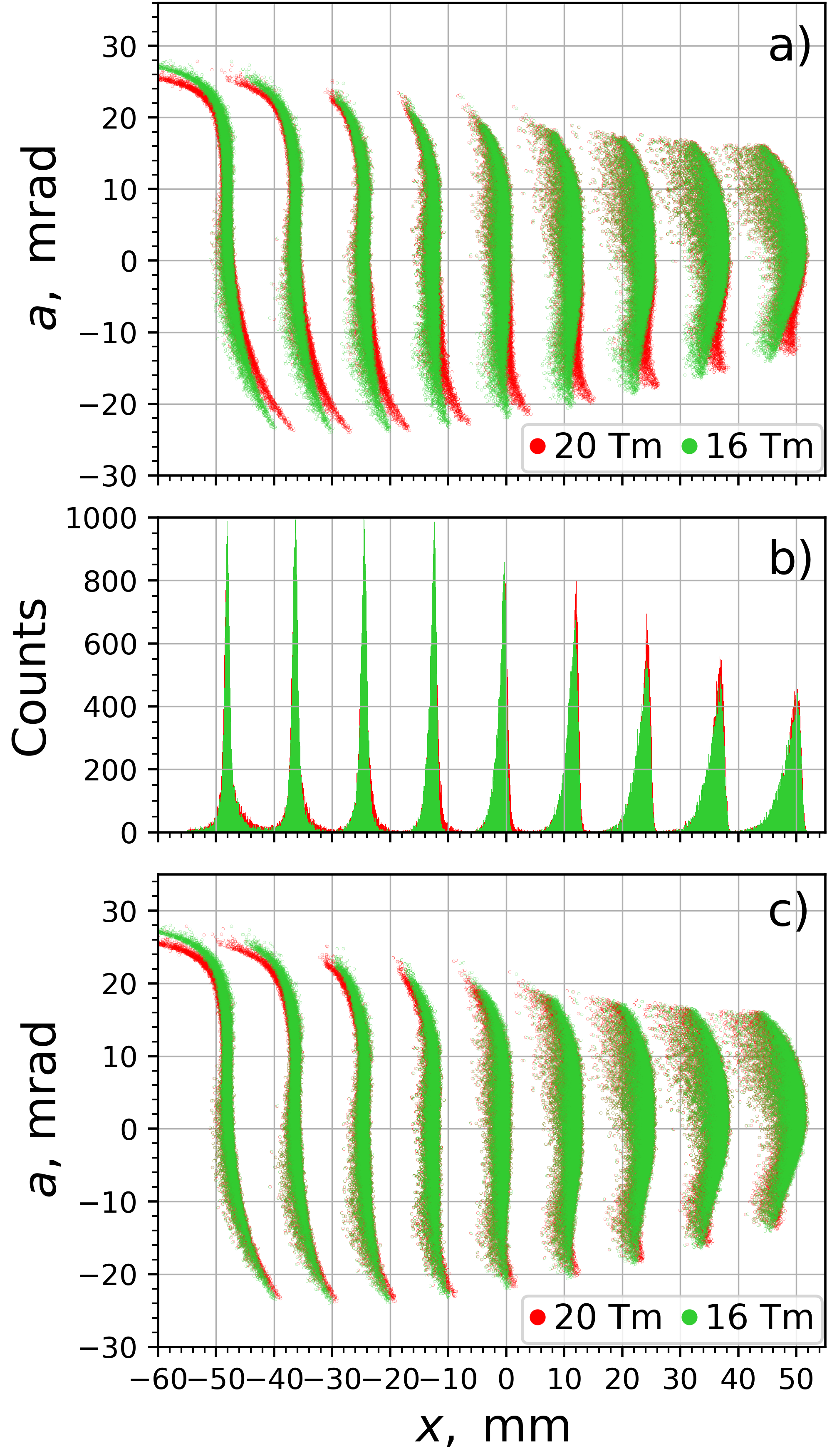}
\caption{The horizontal phase space at the dispersive focal plane after passing through 4 dipole stages in the spectrometer mode for rigidities of 16\,Tm and 20\,Tm. \textbf{a)} the optimal multipole setting for 16 Tm is taken for both cases. \textbf{b)} The number of counts along the $x$-axis for the phase space. \textbf{c)} Individual optimal multipole settings were used.}
\label{Fig:4timesPresepRes}
\end{figure}

Besides separation mode, the SFRS can be used as a high-resolution spectrometer. In this mode, the dispersions of many stages are added. We have simulated such a case with 4 stages to see the effect of saturation on resolution. For this simulation we repeated the first stage of the SFRS preseparators 4 times. The optimal multipole settings for 16 Tm were used for all rigidities. To distinguish the saturation-caused aberrations we artificially compensated magnification at all stages except for the last stage. In Fig. \ref{Fig:4timesPresepRes} \textbf{a)} the resulting horizontal phase space is shown for 9 monoenergetic slices, evenly distributed within $\Delta p/p = \pm 4.8\times 10^{-3}$, and having the same initial distributions in geometrical phase volume. The largest deviation occurs between 16 Tm (green dots) and 20 Tm (red dots), whereas the difference between the distributions from 2 to 16 Tm is relatively small. 
The histogram in Fig. \ref{Fig:4timesPresepRes} \textbf{b)} reveals a slight brightening of the peaks introduced by the non-compensated saturation.

If the optimal settings for each rigidity is used, the difference practically vanishes as shown in Fig \ref{Fig:4timesPresepRes} \textbf{c)}.

To conclude, the saturation in the dipole magnets does not have significant impact on the resolution of the Super-FRS.
On the other hand, taking the higher orders into account is crucial if the design geometrical acceptance of the Super-FRS is going to be used.

\section{Comparison of 3D and MS+Enge FF maps.}
\label{DiffAppr}
For a comparison between 3D and MS+Enge FF maps, we inserted both into the SFRS preseparator optics and studied the differences in phase space distribution in the focal planes and in the optimal multipole settings.

\begin{figure}[t!]
\includegraphics[width=1\textwidth]{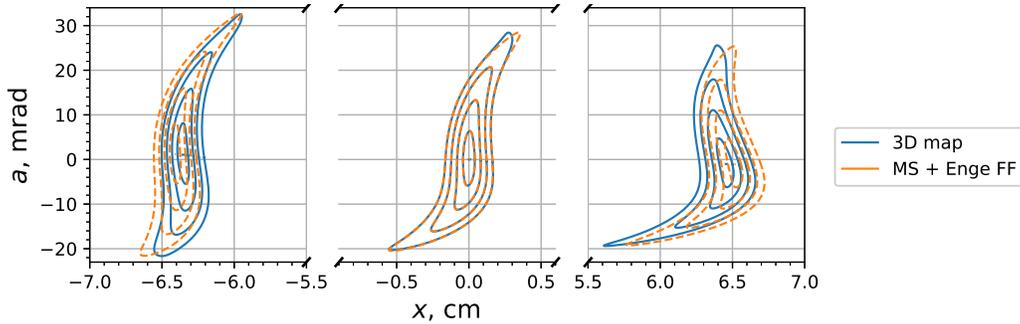}
\caption{The horizontal phase space at FPF2 using 3D map v.s. MS+Enge FF. In both cases the optimal setting for 3D maps on 16 Tm were used. Left, middle and right spot positions correspond to $\Delta p/p$ equal +2.5\%, 0 and -2.5\%, respectively.}
\label{Fig:PRESEP_3DvsMS10}
\end{figure}

In Fig. \ref{Fig:PRESEP_3DvsMS10} the horizontal phase space at the dispersive focal plane FPF2 is compared for both approaches for the particles with $\Delta p/p=$-2.5\% (right), 0 (middle) and +2.5\% (left) and initial coordinates distributed over 4 concentric ellipses \[x_{i}\in \{0.25x_{\textrm{max}},0.5x_{\textrm{max}},0.75x_{\textrm{max}},x_{\textrm{max}}\}\] and \[a_{i}\in \{0.25a_{\textrm{max}},0.5a_{\textrm{max}},0.75a_{\textrm{max}},a_{\textrm{max}}\}.\] In both cases, the optimal setting for 3D maps on 16 Tm were used. For $\Delta p/p=0$ a difference in $x$ of about 1\,mm is observable for $a_{\textrm{max}}$. For $\Delta p/p=\pm2.5$\% the main effect is the shifting of the flanks at about 0.5\,mm towards outside for MS+Enge FF, and this shift is insignificant in comparison to the beam spot size.

\begin{figure}[t!]
\includegraphics[width=0.5\textwidth]{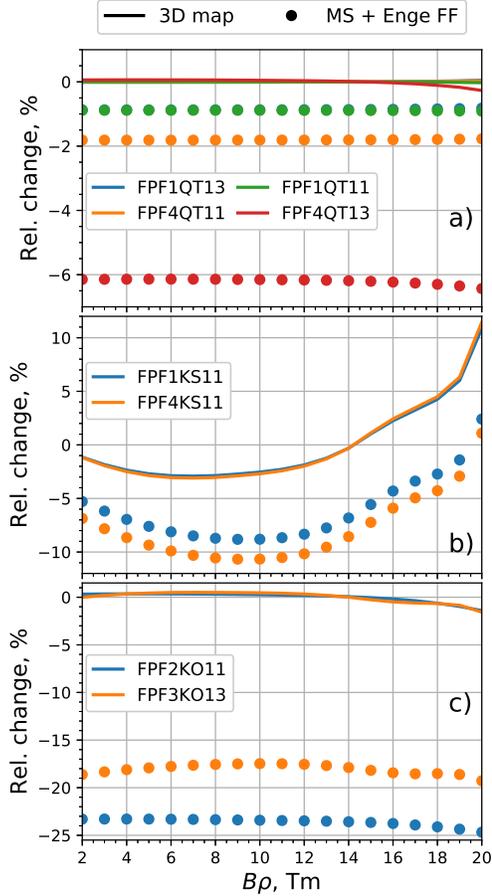}
\caption{Dependence of the relative optimal multipole strengths on the particle rigidity $B\rho$ for the four quadrupoles \textbf{a)}, two sextupoles \textbf{b)}, and two octupoles \textbf{c)}, by comparing 3D maps and MS+Enge FF.}
\label{Fig:PRESEPRelChQSO_MS12}
\end{figure}

In Fig. \ref{Fig:PRESEPRelChQSO_MS12} we show the normalized relative multipole strength changes for four quadrupoles, two sextupoles, and two octupoles. These curves are representative and demonstrate excellent agreement in the shape of the optimal settings for the most multipoles.
The deviations in absolute values arise from the inequality in the lower order terms for 3D and MS+Enge FF, which results in different optimal multipole settings. 

This indicates that the MS+Enge FF approach is valid to find good operation settings quickly.
The deviation in the transverse horizontal phase space distributions between the two methods is very small for the FPF2. Nevertheless, the entire SFRS is about 7 times longer and a larger difference for the quadrupoles is expected.
\section{Conclusion and outlook}
\label{S:conclusion}
We have developed a universal approach for the computation of high order Taylor transfer maps with rigidity dependence and applied it to the Super-FRS-preseparator dipole magnet. The effects of higher orders and magnetic saturation on the images in the separation and in the high dispersion modes were studied. The saturation effects of the dipole magnets occur primarily beyond the rigidity of 16 Tm and and can be well compensated by tuning available multipoles. 

Including higher order terms up to 12\textsuperscript{th} is required for the large acceptance SRFS dipole magnets. 
The saturation effects should be studied further for the Super-FRS quadrupoles as their pole tip field increases from 0.4 up to 4 T. 
Thus we expect much larger changes in transfer maps than for the case of the dipole with maximal field 1.6 T.

Having the rigidity dependent transfer maps together with the measured integral excitation curves for all types of the Super-FRS magnets will allow us to build a precise ion-optical model. This model will make it possible to predict all possible aberrations and optimize the machine performance for arbitrary beam parameters.

\section{Acknowledgement}
E.K. is grateful for support by Franz Klos and Hanno Leibrock regarding magnetic measurements and modeling. 

\bibliographystyle{model1-num-names}

\end{document}